# A PICARD NEWTON METHOD TO SOLVE NON LINEAR AIRFLOW NETWORKS

Harry BOYER, Alain BASTIDE, Philippe LAURET, Franck LUCAS
Université de La Réunion – Laboratoire de Génie Industriel
Equipe Génie Civil Thermique de l'Habitat, IUT de Saint Pierre
40 avenue de Soweto 97410 Saint Pierre
Ile de La Réunion – France
harry.boyer@univ-reunion.fr

**ABSTRACT**

In detailled buiding simulation models, airflow modelling and solving are still open and crucial problems, specially in the case of open buildings as encountered in tropical climates. As a consequence, wind speed conditioning indoor thermal comfort or energy needs in case of air conditionning are uneasy to predict. A first part of the problem is the lack of reliable and usable large opening elementary modelling and another one concerns the numerical solving of airflow network. This non linear pressure system is solved by numerous methods mainly based on Newton Raphson (NR) method. This paper is adressing this part of the difficulty, in our software CODYRUN. After model checks, we propose to use Picard method (known also as fixed point) to initialise zone pressures. A linear system (extracted from the non linear set of equations) is solved around 10 times at each time step and NR uses this result for initial values. Known to be uniformly but slowly convergent, this method appears to be really powerful for the building pressure system. The comparison of the methods in terms of number of iterations is illustrated using a real test case experiment.

## 1. Introduction on multizone airflow network modelling

In a building, wind, thermal buoyancy and ventilation systems combine to create an airflow distribution. The reference pressure of rooms being unknown, the solving of the air weight balance leads to the airflow rates determination. For multizone buildings, only numerical solutions are reachable, in case of successful numerical solving. An analogic network representing the problem can be drawn. Each of the reference pressures of the zones as well as the outside pressure correspond to a node in the network. Conductances linked to the wind or to thermal buoyancy are placed between the pressions. For the simple building (taken to comprise only small openings) in the following figure, the corresponding analogic network is associated :

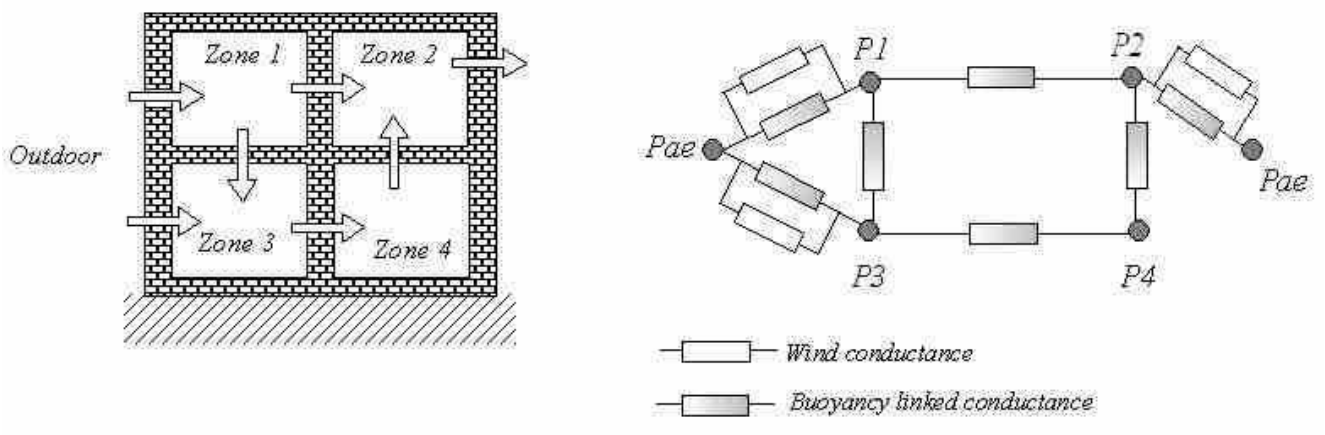

*Figure 1 : A pressure network*

These aspects are fully developped in many textbooks or publications [1-3], and are not developped here. An important notice is that most of the published material deals with small openings, in which flow in unidirectional and respond to well known Crack Flow equation, as $\dot{m} = K(\Delta P)^n$, *K* being link to the permeability of the aperture and *n* the fractional exponent (typically 2/3). Mass balance of each zone (with mechanical ventilation) leads to a non linear system, the unknown being the reference pressures of each zone.

$$\begin{cases} \sum_{i=0, i\neq 1}^{i=N} \dot{m}(i,1) + \dot{m}_{vmc}(1) = 0 \\ \sum_{i=0, i\neq 2}^{i=N} \dot{m}(i,2) + \dot{m}_{vmc}(2) = 0 \\ \quad \ldots \\ \sum_{i=0}^{i=N-1} \dot{m}(i,N) + \dot{m}_{vmc}(N) = 0 \end{cases}$$

that we will note $f(p^n) = B(p)$

In this general form, $\dot{m}(i,j)$ is the airflow rate (kg/s) of air from zone *i* to *j*, $\dot{m}_{vmc}(k)$ the airflow extracted by mechanical ventilation and *N* the total number of zones.

At each time step, the convergence of the obtained system is known to be uneasy to reach. This is linked to mathematical properties of the system to solve and to the large variability of sollicitations (wind and thermal buoyancy) during the hourly time step often used by simulation codes. In many realistic cases, with rigourous convergence criteria, convergence of the solver is not obvious and a quite large amount of iterations is necessary.

## 2. Review on encountered methods

A review on airflow codes shows that the widely used method is Newtown Raphson (NR), often completed with improvements. This method, detailled in [4], leads to the following matricial equation :

$$\mathbf{p}^{n+1} = \mathbf{p}^n - \frac{f(\mathbf{p}^n)}{J(\mathbf{p}^n)}$$, with

$\mathbf{p}^n$ previous time step vector of pressures

$\mathbf{p}^{n+1}$ unknown pressures

$J(\mathbf{p}^n)$ Jacobian matrix

The previous equation can also be written $J(\mathbf{p}^n)\boldsymbol{D}^n = -\mathbf{f}(\mathbf{p}^n)$

$\boldsymbol{D}^n = (\mathbf{p}^{n+1} - \mathbf{p}^n)$ is the corrective terms vector. A consequence of the truncation of term with order more than one in the developement $\mathbf{f}(\mathbf{p}^n)$ to obtain the starting equation is that the pressures found when solving the linear system are not the final solutions to our problem. These values are only an approximation and therefore an iterative method is needed to reach the desired solution. The use of this method requires further explanation for its application to our pressure system.

In order to promote convergence, numerous, different and sometimes combined strategies (or recipes) are usually encountered in litterature. These are for example linked to speed up convergence by relaxation coefficient choice (fixed, variable, optimised, ...). Other authors put their effort on realistic initial values to increase convergence. In one major software, ESP, in it version described in [5], problem geometric description is iteratively modified (in term of diminution of large opening size) to find an intermediate solution used to initialise the solving of another system closest to the one to solve.

Thus the method leads only to quadratical convergence when the estimate is close to the solution. In our case, the physical analysis of the problem leads to consider the evolution of the pressures as a succession of steady states.

In the majority of publications relative to airflow systems solving methods the previous time lapse pressure vector is used for initialising the iterative procedure. In the case of important pressure variations between two different time periods (due to the wind or imposed airflows from mechanical ventilation), it can arise that the previous time lapse pressure vector is outside the convergence field of the numerical method. Walton therefore puts forward a method of pressure vector initialisation by linearising all the airflow equations (the airflow exponent is taken as equal to 1), so the initial pressure vector considered is the solution to a linear system which characterises the laminar state in the building.

Specific problems arise when taking into account large openings. For vertical internal large openings, separating zones, two methods are encountered. One is based on Bernouilli's equation and leads to the speed field integration, after calculation of the neutral height. The one we have choose is Walton model [6], leading to splitting up of large openings in two small openings. For these two equivalent small openings, the model considers specific heights (5/18 and 13/18 of large opening vertical size), exponents (0.5) and discharge coefficient (0.78). It is therefore possible to couple large openings to the previous obtained non linear system. To complete the review, we must bear in mind that there are not many papers that have been published on horizontal openings and that only a few methods are available for external openings. With the consideration of large openings, many convergence problems appear and lead us to propose an improvment of initialising pressures.

Furthermore, the integration of large openings into a pressure system can cause problems of convergence speed. Located between a zone and the outside, a large opening links strongly the inside and the wind pressures. As in between two time lapses, the wind speed and direction can change considerably, the remarks of the previous paragraph apply. Another source of problems is the value of 0.5 found for the airflow exponent of the equivalent small openings in Walton's model. If the mass balance is symmetrical in relation to the pressures and that the exponents are equal to 0.5, the Newton method diverges irremediably. Feustel insists that the convergence of the method lowers as the number of exponents equals 0.5 grows [2].

These divergence problems being ignored for the moment, the large openings may also compromise the speed of the convergence. It is furthermore established that a small pressure difference generates important mass flows, through a large opening. For the zones considered, the mass balances partial derivatives have important numerical values (compared to a case concerning only small openings). Consequently, the amplitude of the successive corrective terms is low. In these conditions, an important number of iterations is necessary to reach the solution. For a building which comprises various zones, separated by large openings, various directions exist in which the convergence is slow. Between two time lapses, the distance between the pressure vectors is a function of the disturbance gradient due to the solicitations. The number of iterations can therefore be very important and also change considerably from one time period to another.

Various techniques have been programmed in CODYRUN, including NR with a systematic under relaxation coefficient value of 0.1, Walton's optimised relaxation coefficient [7] also described in [8], Clarke's method [5] (embedded in earlier version of ESP). In the figures on page 5, we will refer the first method as NR and the second one as WM (as Walton Modified). It is to be noticed that this last method is integrated in major airflow models as AIRNET, COMIS and last version of ESP.

## 3. Elements of checkings of the initial model

The objectives of this part is to ensure that, before the improved method to be implemented, the code give accurate outputs with some intermodel comparison (concerning small openings) and analytical case for consideration of internal vertical large opening.

**AIVC TN 51 test case**

In [9], a test case composed of a building with three storeys is described. All boundary conditions are imposed (wind, external temperature) and indoor conditions are constant (steady state).

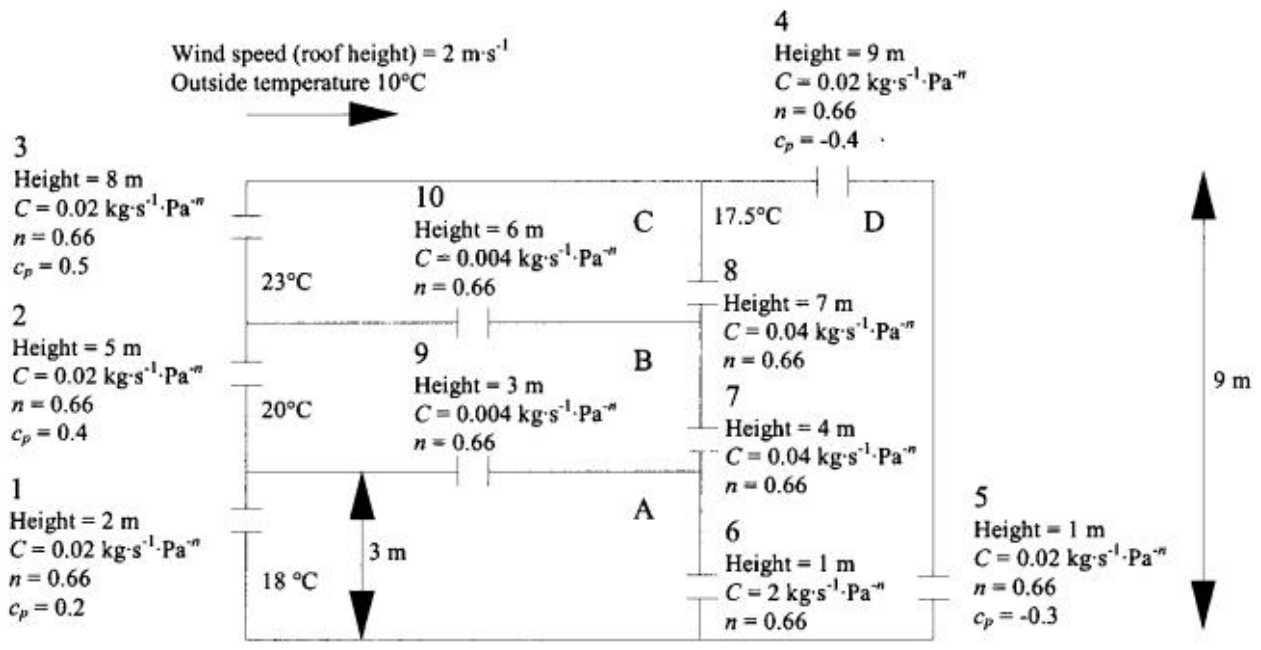

*Figure 2 : TN 51 Test case*

Next figure gives shows solution of 3 references codes, i.e. COMIS, CONTAM93 and BREEZE. CODYRUN's results were added on the figure extracted from the note.

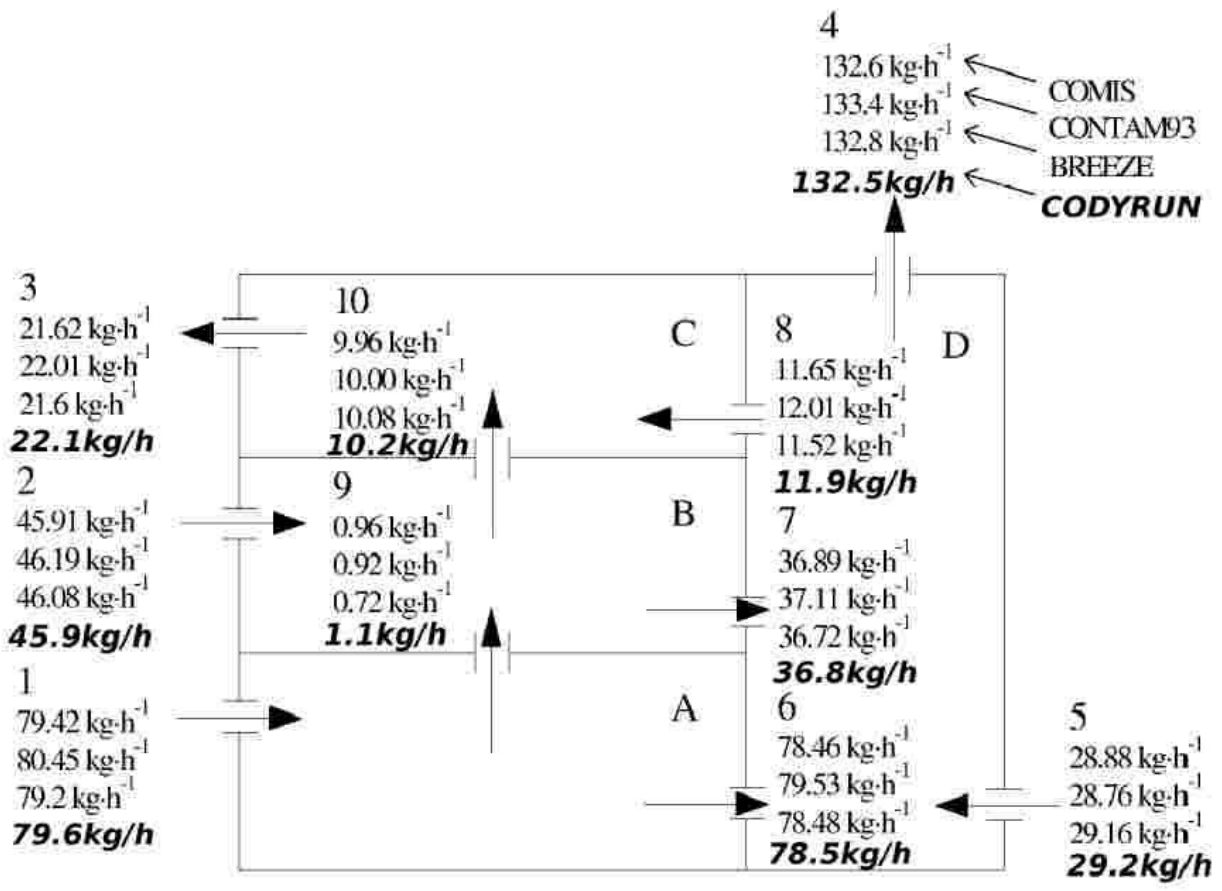

*Figure 3 : Intermodel comparison*

As it can be seen, in terms of numerical results, CODYRUN give nearly the same values as the other codes, little differences being linked to numerical aspects as algorithms or convergence criterias used.

**IEA Task 34**

Another case [10] concerns large openings taking into account and respond to the following sketch.

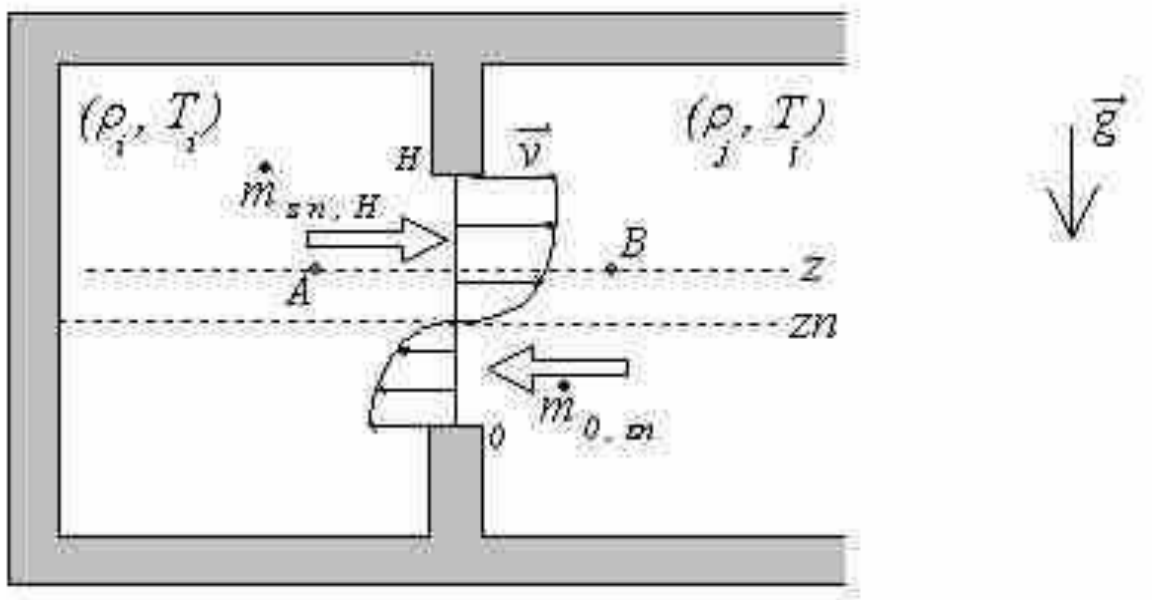

*Figure 4 : IEA Task 34 case*

For this case, analytical expressions can be found using the mass transfer approach :

$$\dot{m} \cong 0.04 W H^{3/2} (\Delta T)^{1/2}$$

Using the possibility to impose temperatures, good agrement is found between the results. For example, with 100 K zone temperature difference, 0.34 $kg.s^{-1}$ are found versus 0.4 $kg.s^{-1}$ for the approximated analytical solution (with $W = H = 1m$)

It is important to notice that the convergence of this case was found to be very sensitive to the convergence criterion used and to the relaxation coefficient choice (if NR with fixed relasation value is used). Other airflow tests cases were performed with success, but won't be reported here.

## 4. The PICARD method

The basic idea is to couple a first order method (low convergence speed but large convergence disk) to guide the numerical scheme close to the solution and the second order NR to reach quickly convergence.

The first order method choose is Picard method, because widely used in CFD. This method is cited by Koldiz [11].

The previous system $f(p^n) = B(p)$ can be rewriten under the form

$$\left[A\left(p^{n-1}\right)\right] p = B(p)$$

and solved iteratively (*k* being the iteration index)

$$\left[A\left(p_k{}^{n-1}\right)\right] p_{k+1} = B(p_k)$$

This problem leads to solution of the linear system, with usual methods. Usual values for the number of iterations is 10. Because of non convergence risk, specially in case of large openings, we prefer to promote convergence using an acceleration factor *a* (0.5). If $p^*$ is the solution of previous linear system, then

$$p_{k+1} = a\ p_k + (1-a)\ p^*$$

No details will be given concerning the computer implementation the filling up of A matrix and B vector. It is to be noticed that in some cases, A matrix can be (or become during iterations) singular (or ill conditionned) and this will have to be detected to avoid next iteration of Picard and give hand to NR (or WM). Identified cases are those with reciprocical exchanges, in which the Picard method is unusable. It was observed that in all the small opening cases, NR and WM becomes completely useless because the values found by the Picard method are very close to the solution.

Picard being used before a non linear solver (NR or WM), we will refer to PNR and PWM in order to indicate its use.

## 5. Illustration with a real case

After modifications, same results were obtained with TN51 and IEA Task 34 and no conclusions can be made on the numerical speed or efficiency improvment, these two cases being in steady state (i.e. pressures no longer vary after the first time step convergence reached).

To obtain a dynamic case, we compare the methods (NR, PNR, WM and PWM) using a dwelling, modelled as 5 zones building with external small openings, two zones being separated by a large opening (sliding door between living room and bedroom 2) and a measured meteorological file. This instrumentation was part of technical evaluation of building prescriptions for French overseas territories. The dwelling, represented on next figure includes three bedrooms and a living room. It is situated under the roof.

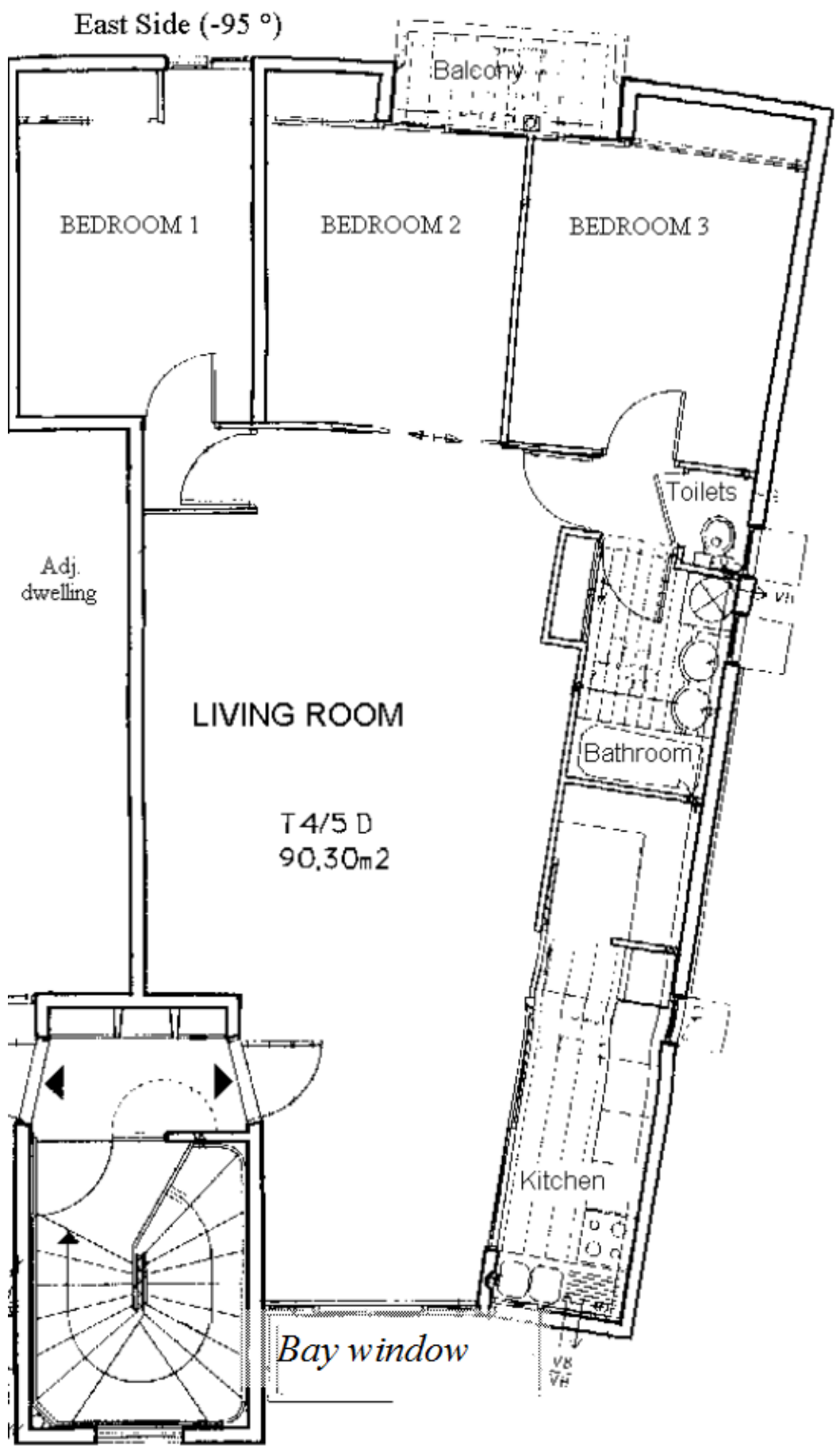

*Figure 5 : La Trinité dwelling*

Simulations were conducted on the measured 10 first days of meteorological file of Saint-Denis, Ile de La Réunion, 1998, using NR, PNR, WM and PWM numerical methods, with the same convergence criteria based on mass balance of each zone (10-3 $kg.s^{-1}$). For each time step (30 mn), the number of iteration is saved in the result file (containing also temperatures, airflow rates, ...).

The comparison of the number of required iterations is shown on the two next figures, the first one on the whole simulation period, the second restricted to the first day for more clarityA first comment is that methods using WM (Walton Modified) are much more efficient than NR01. Concerning PNR, for 141 times step (on a total of 480), the solution is reached after one NR iteration. This is meaning that in 29 % of cases of our simulation, Picard found the solution in at less than 10 iterations and NR is not any more needed. An interesting point is that in a few cases (time step 350 and 422), Picard method leads to a greater number of iterations, which is exatly the contrary of the desired result. This appears to be linked to ill-condition of the Picard linear system. To takle this problem, it appears necessary to truncate the pressure evolution during Picard's iterations (itentionnaly, a too large troncation value of 60 Pa was let).

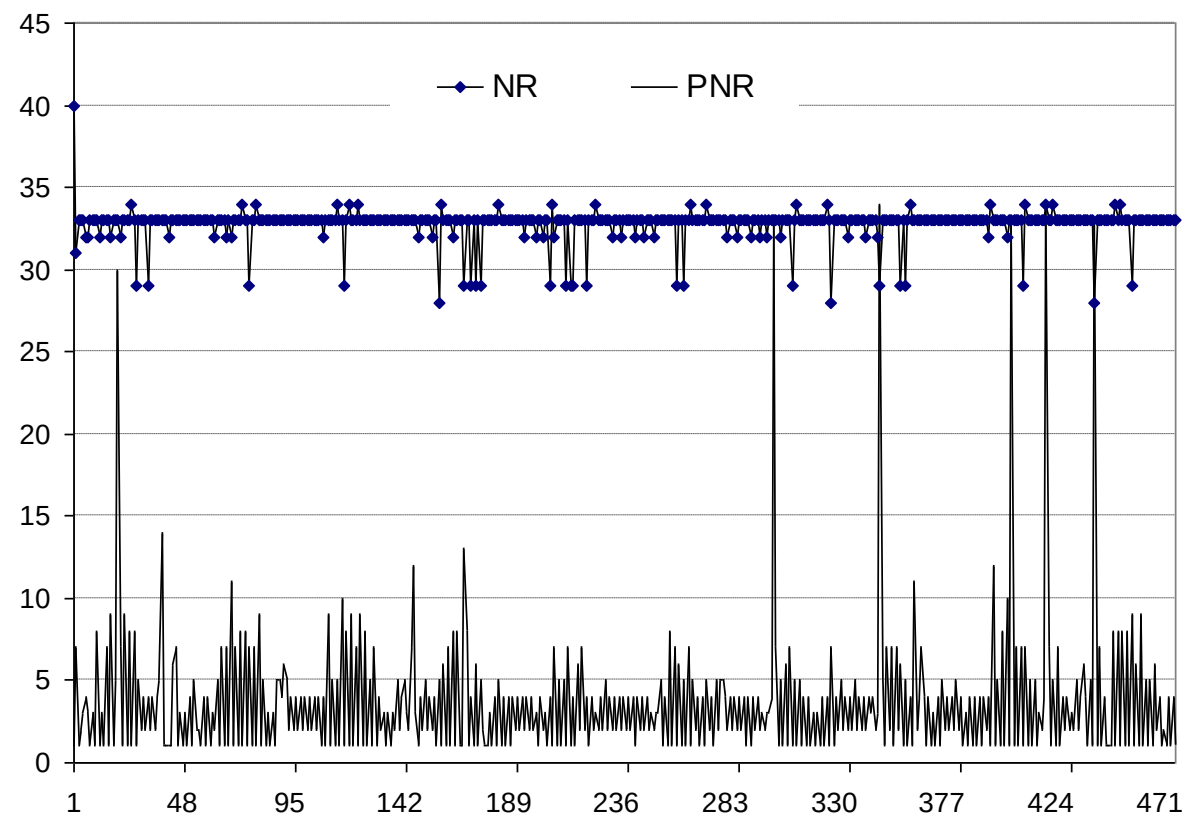

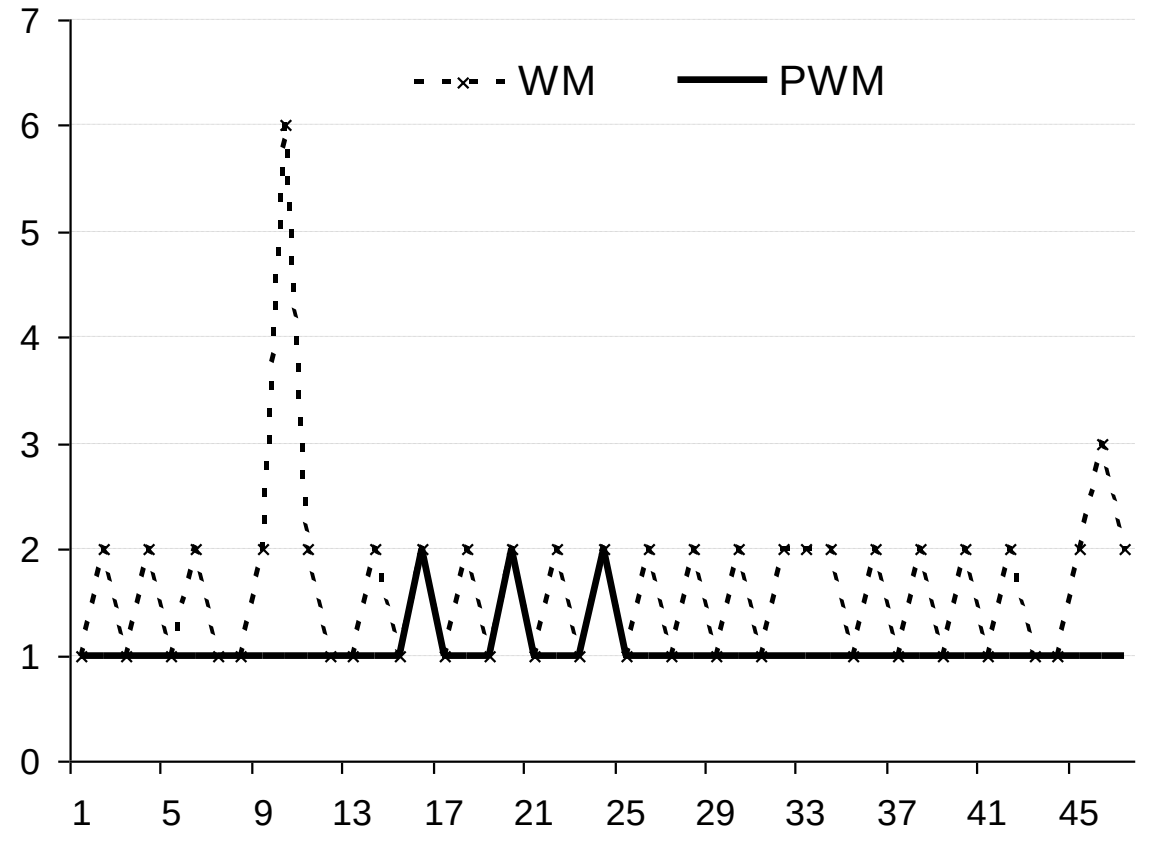

*Figure 6 : Number of required iterations*

With PWM, only a slight improvment is observed, WM leading to low number of iterations. The average number *n* of required iterations is given in the following table, concerning the whole 10 days simulation period.

| | NR | PNR | WM | PWM |
|---|---|---|---|---|
| n | *33* | *4* | *2* | *1* |

Picard's method appears to improve convergence. A reduction of more than eight times is observed for NR and about two for WM and back up the Picard method to initialise pressures in this non linear system solution.

In reality, some CPU time was consumed in Picard's iterations. In case Picard did not find the solution, it is necessary to add 10 to *n* (because of the 10 iterations of Picard). Meanwhile, other simulations (with different buildings, convergence criteria, ...) confirm this improvment and, the more important, no case of non convergence appears till this method was included in the computer software.

## 6) Conclusion

This paper aims at presenting a way of improvement for solution of non linear pressure systems obtained with nodal networks linked to airflow calculations. Although several other methods are encountered, this Picard method appears to be an interesting complement, in particular when considering large openings. The couple Picard and Walton Modified method appears also to secure convergence of the numerical solver. More in details, other cases studies have to be examined (with different size of openings, large external openings, values of parameters as the acceleration factor, troncation value, number of Picard iterations, ...) in order to reach more complete informations about the improvements obtained.